\documentclass[twocolumn,showpacs,showkeys,preprintnumbers,amsmath,amssymb]{revtex4}

\usepackage{graphicx}
\usepackage{dcolumn}
\usepackage{bm}
\usepackage{url}
\usepackage{rotating}
\usepackage{natbib}
\usepackage[font=footnotesize,caption=false]{subfig}

\begin{document}

\title{Migration and Trade: \\ A Complex-Network Approach}

\author{Giorgio Fagiolo}%
\affiliation{LEM, Sant'Anna School of Advanced Studies, Pisa (Italy)}
\author{Marina Mastrorillo}
\affiliation{Princeton University, Princeton (USA)}%

\date{\today}

\begin{abstract}
This paper explores the relationships between migration and trade using a complex-network approach. We show that: (i) both weighted and binary versions of the networks of international migration and trade are strongly correlated; (ii) such correlations can be mostly explained by country economic/demographic size and geographical distance; (iii) pairs of countries that are more central in the international-migration network trade more.
\end{abstract}



\maketitle

\hyphenation{op-tical net-works semi-conduc-tor}

\section{Introduction}
In the recent years, the empirical study of macroeconomic networks has received an increasing attention in the literature \cite{ScienceNets2009}. Macroeconomic networks are graphs where nodes are world countries and links represent interaction channels between countries, concerning e.g. trade \cite{Se03, Ga04,Fa08,Fa09,Fa10}, finance and investment \cite{Song2009,finnet_qf_2010}, banking \cite{minoiu_reyes_2013,contagion_survey} and migration \cite{FagMas13,Davis_etal_2013}. It has been argued that knowledge of the topological properties of these networks may be important to understand how economic shocks propagate and how well countries perform over time \cite{lee_etal_plosone_2010,postmortem_jedc}.

So far, with some minor exceptions \cite{finnet_qf_2010}, such networks have been studied in isolation, as they were distinct interaction layers independently and separately connecting world countries. In reality, however, they represent economic phenomena that are likely to be causally linked through complicated relationships. In order to start addressing the issue of how these different layers are related, one might think to superimpose them one on each other, thus forming a multi-graph, where between any two countries there may exist many links, each representing a different type of between-country interaction (trade, migration, finance, etc.). 

Within this framework, one may be interested in understanding the extent to which different layers display similar topological properties and whether such properties are correlated, or causally linked, between layers. A natural example here concerns the relationship between migration and trade, which has been explored by a huge literature in economics (see for instance \cite{GastonNelson_2011,Egger_etal_2012}). Empirical studies on bilateral trade find quite a robust evidence that trade between any two countries increases the larger the stock of immigrants born in one country and living in the other (bilateral-migration effect) \cite{Parsons_2012}. The reason is found in both the emergence of new consumption-preference patterns and the decrease in transaction costs, due to the better knowledge that migrants have of both home and host country cultural, economic and institutional environments \cite{Gould_1994}. More generally, however, countries are embedded in a complex web of migration channels \cite{Le96}. Therefore, one may argue that, in addition to bilateral migration, also the relative positions of any two countries in such a complex web may affect their bilateral trade.

The paper exploits these ideas and performs two sets of exercises. First, we systematically explore the relationships between the networks of international migration and trade. Recently, a large body of contributions have been separately investigating trade and migration from a complex-network perspective. The topological properties of the International-Trade Network (ITN)  \cite{Se03, Ga04,Fa08,Fa09,Fa10} and the International-Migration Network (IMN) \cite{FagMas13,Davis_etal_2013}, and their evolution over time, have been indeed extensively studied with both a binary and weighted network approach \cite{Ga05,Fa08,FagMas13}. Furthermore, their community structure \cite{Barigozzi_etal_2010physa,Piccardi_Tajoli_2012,FagMas13,Davis_etal_2013} have been separately identified. Finally, both statistical null-network models \cite{Squartini_etal_2011a_pre,Squartini_etal_2011b_pre,FagMas13} and economics-inspired gravity-like models have been fitted to the data \cite{Fagiolo2010jeic,Duenas_Fagiolo_2013,Ward_etal_Gravity_Rainbow_2013} in order to understand the determinants of the observed network regularities in each network. Here, we compare the topological structure of the IMN and ITN and study their correlation patterns. We also investigate the main determinants of these correlations and we find that economic and demographic country size, as well as geographical distance, play a key role in explaining differences and similarities between IMN-ITN topologies.

Second, we ask whether the position of countries in the IMN explains, in addition to bilateral-migration effects, their bilateral trade. We expand upon the existing literature by fitting gravity models of trade where country centrality is added as a further explanatory factor. We find that pairs of countries that are more central in the IMN (e.g. hold more inward migration corridors) also trade more. We interpret this in terms of a third-country effect: the more a pair of countries is central in the IMN, the more they share immigrants coming from the same third-country, and the stronger the impact of forces related to consumption preferences and transaction-cost reduction \cite{Rau99,Fel10,Fel12}. Interestingly, we find that also inward third-party migration corridors that are not shared by the two countries can be trade enhancing, in addition to common inward ones. We suggest that this can be due to either learning processes of new consumption preferences by migrants whose origins are not shared by the two countries (e.g. facilitated by an open and cosmopolitan environment) or by the presence in both countries of second-generation migrants belonging to the same ethnic group.       

The rest of the paper is organized as follows. Section \ref{data} describes the data and defines the International Migration-Trade Network (IMTN). Section \ref{topology} discusses the topological properties of the IMTN. Section \ref{regressions} presents gravity-model estimation results. Section V concludes.

\section{The International Migration-Trade Network: Data and Definitions\label{data}}

Migration data employed in the paper come from the United Nations
Global Migration Database \cite{Ozden_etal_data_2011}, which comprises, for each year $y=\{1960,1970, 1980, 1990, 2000\}$, an origin-destination square matrix recording bilateral migration between 226 countries. The generic element $(i,j)$ of each matrix is equal to the stock of migrants (corresponding to the last completed census round) originating in country $i$ and present in destination $j$, where migrant status is consistently defined in terms of country of birth.

As to trade, we employ the dataset provided by Kristian Gleditsch \cite{GledData2002}, which contains bilateral export-import yearly figures for the period 1950-2000. Trade matrices follow the flow of goods: rows represent exporting countries, whereas columns stand for importing countries. The generic bilateral element $(i,j)$ thus records exports from $i$ to $j$ in a given year. Trade figures, which are originally expressed in current US dollars, are then deflated to get real values.

We merged these two datasets by keeping, in each of the 5 years available in migration data, all countries that were present also in trade data with at least a positive import or export flow. This results in 5 origin-destination $N^y \times N^y$ matrices, where $N^y=\{109, 135, 158, 163, 183\}$. The sample of countries included explains more than 90\% of total world trade flows and migration stocks in each year.

We employ additional country-specific data such as real gross domestic product (rGDP), population (POP) and per-capita real gross-domestic product (rGDPpc) from Penn World Tables version 6.3 ({https://pwt.sas.upenn.edu}). We also use bilateral country geographic, political and socio-economic data from the CEPII gravity dataset (see {http://www.cepii.fr}). The latter includes information about between-country geographical distance ($\delta$)\footnote{We employ the great-circle definition of country distances, see {http://www.cepii.fr/anglaisgraph/bdd/distances.htm}. Results do not change using alternative distance definitions.}, contiguity (CONTIG, i.e. whether two countries share a border), existence of a bilateral trade agreements (RTA), common language (COMLANG), etc. and will be mostly used to perform gravity-like exercises (see Section \ref{regressions}).
 
We use trade and migration data to build a time sequence of 5 weighted-directed migration-trade (multi) graphs describing both bilateral-migration stocks and exports flows. More precisely, we define the international migration-trade network (IMTN) as a directed weighted multigraph wherein between any two nodes (countries) there can be at most four weighted-directed links, two of which describing bilateral export and the other two concerning bilateral migration. Alternatively, we can think to the IMTN as a time sequence of 2-layer weighted directed networks, the first layer representing the IMN and the second the ITN. In both cases, the IMTN at each time $y = 1960, \dots ,2000$ is characterized by the pair of $N^y \times N^y$ weight matrices $(M^y, T^y)$, where $M^y$ and $T^y$ define respectively the weighted-directed International Migration Network (IMN) and the weighted-directed International Trade Network (ITN). The generic element of $M^y$ represents the stock of migrants $m^y_{i,j}$ originated in country $i$ and present at year $y$ in country $j$. Instead, the generic element of $T^y$ records the value of exports $t^y_{i,j}$ from country $i$ to country $j$ in year $y$.

Accordingly, we define the binary projection of the IMTN through the pair of $N^y \times N^y$ adjacency matrices $(A_M^{y},A_T^{y})$, where the generic element of $A_X^{y}$, $X=\{M,T\}$, is equal to one if and only if the correspondent entry in $X^y$ is strictly positive (and zero otherwise).   


\begin{figure}[h]
\centering
\subfloat[]{\label{fig:instrength}\includegraphics[width=3.6in]{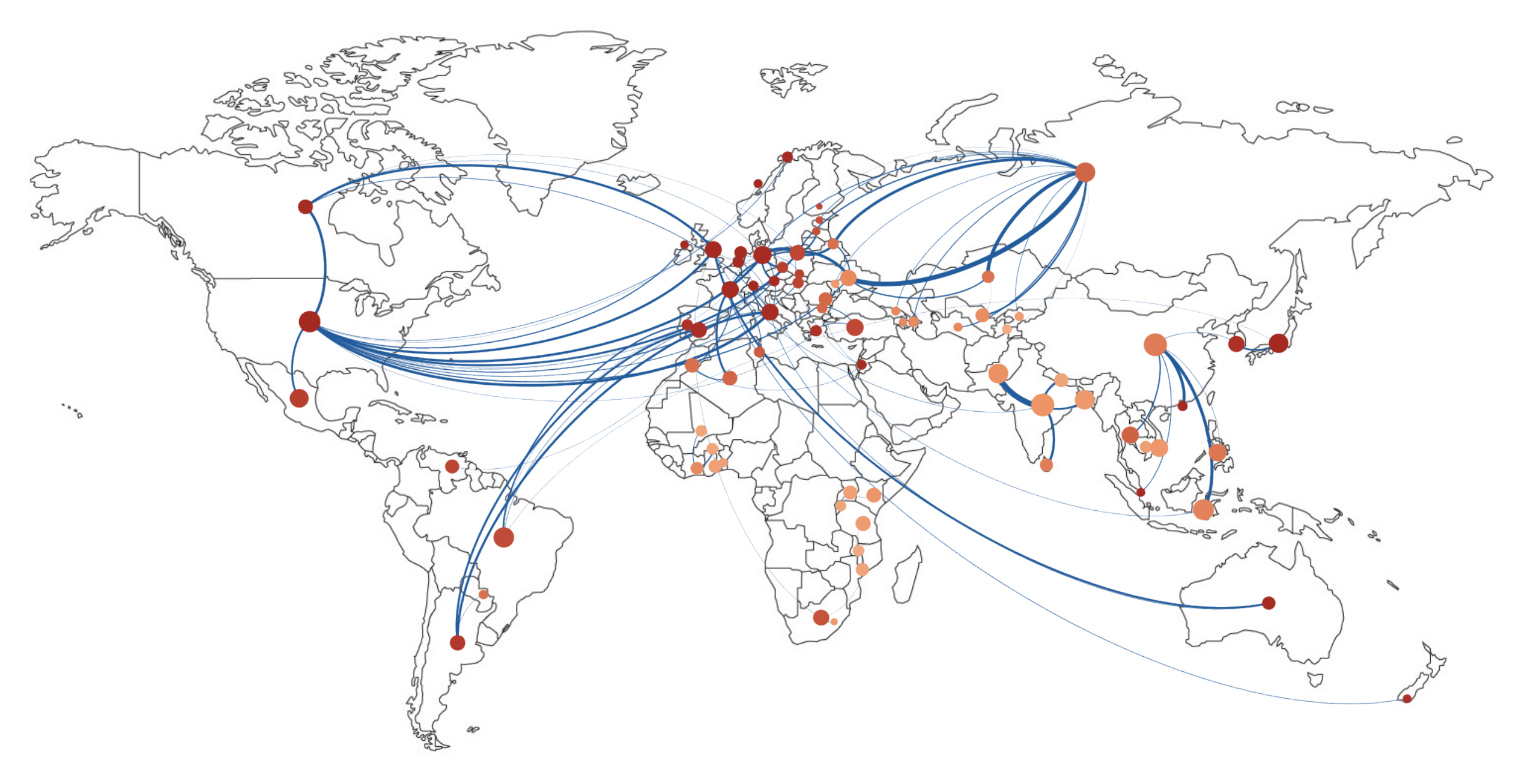}} \\
\subfloat[]{\label{fig:outstrength}\includegraphics[width=3.6in]{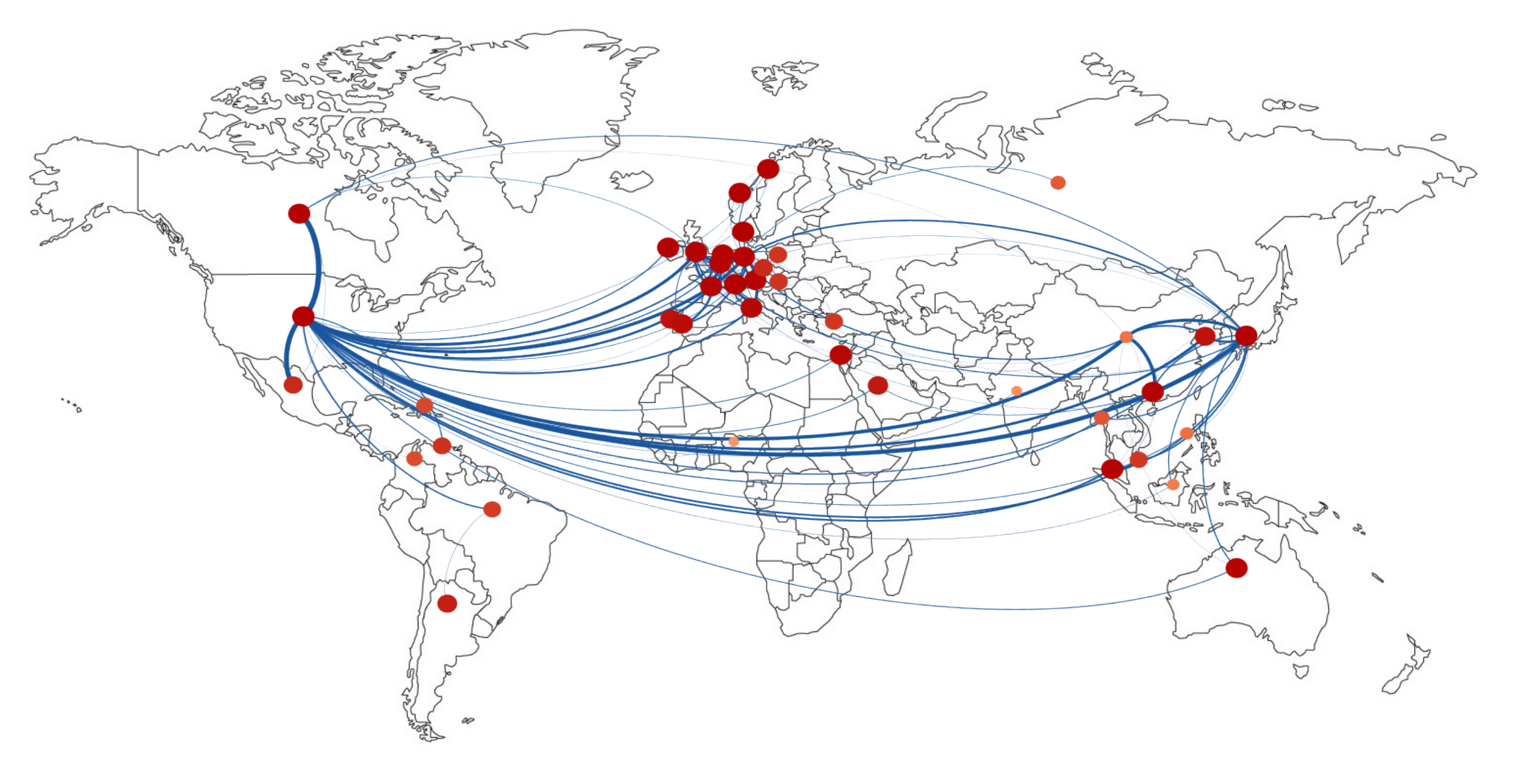}}
\caption{The International-Migration Network (a) and the International-Trade Network (b) in year 2000. The figure plots the undirected weighted version of the ITN and IMN where only top 5\% of bilateral link weights (total number of bilateral trade and total number of bilateral migrants) are drawn. Tickness of links in the plot is proportional to the logs of link weights. Node size is proportional to the log of country population. Node color represents country income (rGDPpc), from beige (low-income countries) to red (high-income countries).}
 \label{IMTN}
\end{figure}

Figure \ref{IMTN} plots the undirected weighted version of the IMN (a) and of the ITN (b) in year 2000. In the figures, link directions are suppressed to attain a better visualization of the graphs and only the top 5\% of link weights are plotted. Link thickness is proportional to total bilateral migrants ($m^y_{i,j} + m^y_{j,i}$) and total bilateral trade ($t^y_{i,j} + t^y_{j,i}$), respectively. To get a feel of migration and trade determinants, node size is made proportional to the log of country population, while node color (from beige to red, i.e., from lighter to darker grey) represents logs of country rGDPpc (a measure of country income). ￼The map allows one to appreciate some of the main general differences between IMN and ITN, e.g. the central role of Russia in the IMN (absent in the ITN) and the strong trade connections between the United States and South-Asian countries (absent in the IMN). Also, as expected, notice the widespread presence of low-income countries in the IMN (beige color), while the most relevant trade connections occur between countries with higher rGDPpc (red color).

\section{Migration vs Trade: A Comparative Network Analysis}\label{topology}

We begin with a comparison between the topological properties of the two IMTN layers across time. Table \ref{Descriptives} reports for the years 1960, 1980 and 2000 the main features of the two networks. Note that both networks are extremely dense. The ITN increased its density by 50\% during the period covered by our data, and became more dense than the IMN in 2000. As expected, the ITN is also more symmetric than the IMN, as testified by a larger bilateral density (i.e. the percentage of reciprocated directed links). This is because a trade channel is more easier to reciprocate than a migration corridor. This is true also when one takes into account the weights of the links: weighted asymmetry (computed as in \cite{Fagiolo2006EcoBull}) is indeed larger in the IMN, capturing the fact that countries tend to be more bilaterally balanced in trade than in migration. Note also that both networks are always weakly and (almost) strongly-connected. Indeed, the number of weakly connected components is always one and strong connectivity is not achieved before year 2000 only because of the presence of one or two (strongly) not-connected countries, typically small and peripheral nations. Finally, as already noticed in Refs. \cite{Fa09,FagMas13}, the IMN features a more marked small-world property, with average-path lengths smaller than in the ITN.  

\begin{table*}[ht]\centering 
\renewcommand{\arraystretch}{1.3}
\label{descriptive_statistics}
\begin{tabular}{l|rr|rr|rr} 

\hline \hline
 &	\multicolumn{2}{c|}{1960}	&	 \multicolumn{2}{c|}{1980}  &\multicolumn{2}{c}{2000} \\ \hline

 &	\multicolumn{1}{c}{ITN}	& \multicolumn{1}{c|}{IMN} & \multicolumn{1}{c}{ITN}	&  \multicolumn{1}{c|}{IMN} & \multicolumn{1}{c}{ITN} &  \multicolumn{1}{c}{IMN} \\ \hline
No. Nodes  & 	109	 & 109 	 & 158	 & 158 & 183 & 	183 \\
Density	 & 0.3843 & 	0.5808  & 0.4628 & 	0.5080 & 	0.5687	 & 0.5503 \\
Bilateral Density & 	0.8439 & 	0.7234  & 0.8697 & 0.6975 &  0.9802 & 	0.7097 \\
Weighted Asymmetry  & 	0.1424 & 	0.1886  & 0.0953 & 	0.5514 & 0.1151 & 0.6615 \\
No. SCC	 & 3	 & 2 &  3	 & 2	 & 	1   & 1 \\
Size Largest SCC & 	107	 & 108 & 	156	 & 157  & 183 & 	183 \\
No. WCC	 & 	1 & 	1	  & 1	 & 1	& 1 & 1 \\
APL (Undirected) & 	1.5646 & 	1.2586  & 1.4811 & 	1.3383  & 1.4217 & 1.2899 \\
\hline \hline
\end{tabular}
\caption{IMN vs. ITN: Descriptive Network Statistics. \textit{Note}: SCC: Strongly connected components. WCC: Weakly connected components. APL: Average path length. \label{Descriptives}}
\end{table*}

\begin{figure}[h!]
\centering
\includegraphics[width=3in]{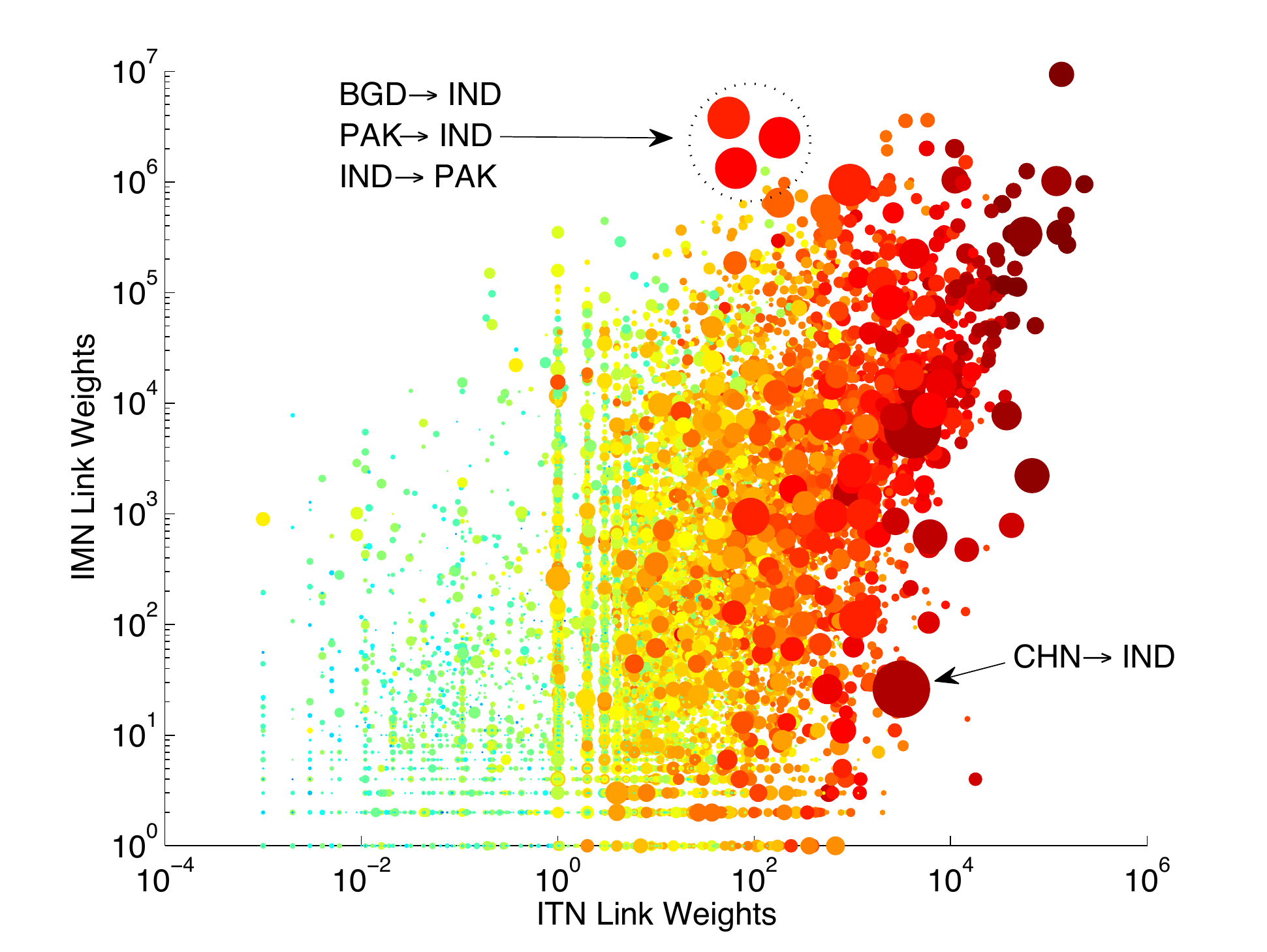}
\caption{IMN vs ITN link weights. Logaritmic scale. Markers size is proportional to the log of POP$_i^y$*POP$_j^y$/$\delta_{ij}$. Colors scale (blue to red) is from lower to higher values of logs of rGDP$_i^y$*rGDP$_j^y$/$\delta_{ij}$. Year=2000.}
\label{weights_scatter_2000}
\end{figure}

We now study the extent to which the two layers of the IMTN display any correlated behavior. We start exploring whether link weights $(m^y_{i,j},t^y_{i,j})$ are positively related, and why. Figure \ref{weights_scatter_2000} shows a scatter of link weights in the ITN vs IMN (log scale) in year 2000. Each dot represents, in the space $(m^y_{i,j},t^y_{i,j})$, an ordered pair of countries $(i,j)$ for which either $m^y_{i,j}>0$ or $t^y_{i,j}>0$. Note first how a stronger link-weight in the ITN is typically associated to a stronger migration link-weight: if $i$ exports a higher trade value to $j$, in $j$ there is also a larger stock of migrants originated in $i$. To explain why, each dot has been giving a size proportional to the product of country population divided by country distance, and a color scale (from blue to red) depending on the product of country rGDPs divided again by geographical distance. The rationale for this exercise lies in the well-known empirical success of the gravity model for both migration and trade \cite{GravityBook,Lewer_2008}, which states that bilateral trade flows (respectively, migration stocks) are well explained by a gravity-like equation involving country sizes (rGDP and POP, respectively) and, inversely, geographical distance. If this is the case, one should expect that most of the variation in the cloud of points $(m^y_{i,j},t^y_{i,j})$ can be explained by larger country sizes and smaller distances. 

This is actually what Figure \ref{weights_scatter_2000} suggests: red and large dots (higher values for POP$_i^y$*POP$_j^y$/$\delta_{ij}$ and rGDP$_i$*rGDP$_j$/$\delta_{ij}$) are located in the north-east part of the plot. This is more evident for the relation between trade, rGDP and $\delta$, than in the case of migration. Indeed, there exist large (and red) dots characterized by high trade values but relatively low migration stocks. This is the case of migration of Chinese people to India, which is historically feeble, unlike correspondent exports flows. Similarly, there are large dots associated with intermediate trade levels and very high migration stocks. These refer to the triangle Bangladesh, India and Pakistan, which experienced huge migration flows at the time of partitioning of India.

\begin{figure}[h]
\centering
\includegraphics[width=3in]{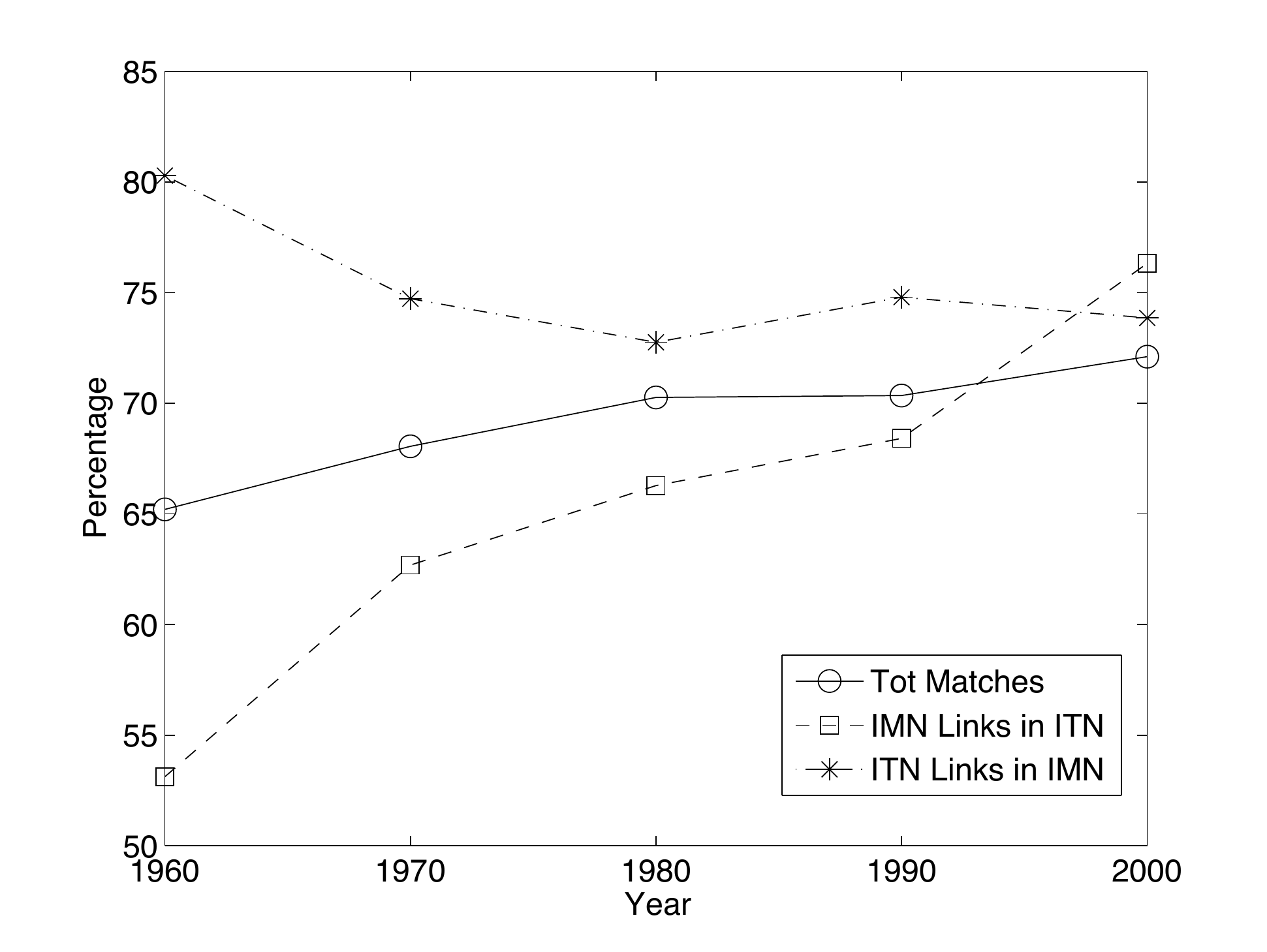}
\caption{IMN vs ITN: Comparison of binary structure. Tot Matches: \% of total matches (either missing or present links). IMN Links in ITN: \% of IMN links which are also present in the ITN. ITN Links in IMN: \% of ITN links which are also present in the IMN.}
\label{bin_percentages}
\end{figure}

We move now to investigating matches and mismatches between ITN vs IMN binary structures. We do so by comparing adjacency matrices $(A_M^{y},A_T^{y})$ and counting the percentage of total matches (either present or missing links), and the share of IMN links (respectively, ITN links) which are also present in the ITN (respectively, in the IMN). Results are presented in Figure \ref{bin_percentages}. Two main findings stand out. First, the two networks have become more and more similar in terms of presence/absence of links. Second, this has happened thanks to an increasing number of migration corridors that became also trade channels. On the contrary, the share of trade channels that are also migration corridors remained constant and even declined. This hints to a causal link from migration to trade, which we shall explore in more details in Section \ref{regressions}.   
 
\begin{figure}[h]
\centering
\includegraphics[width=3.2in]{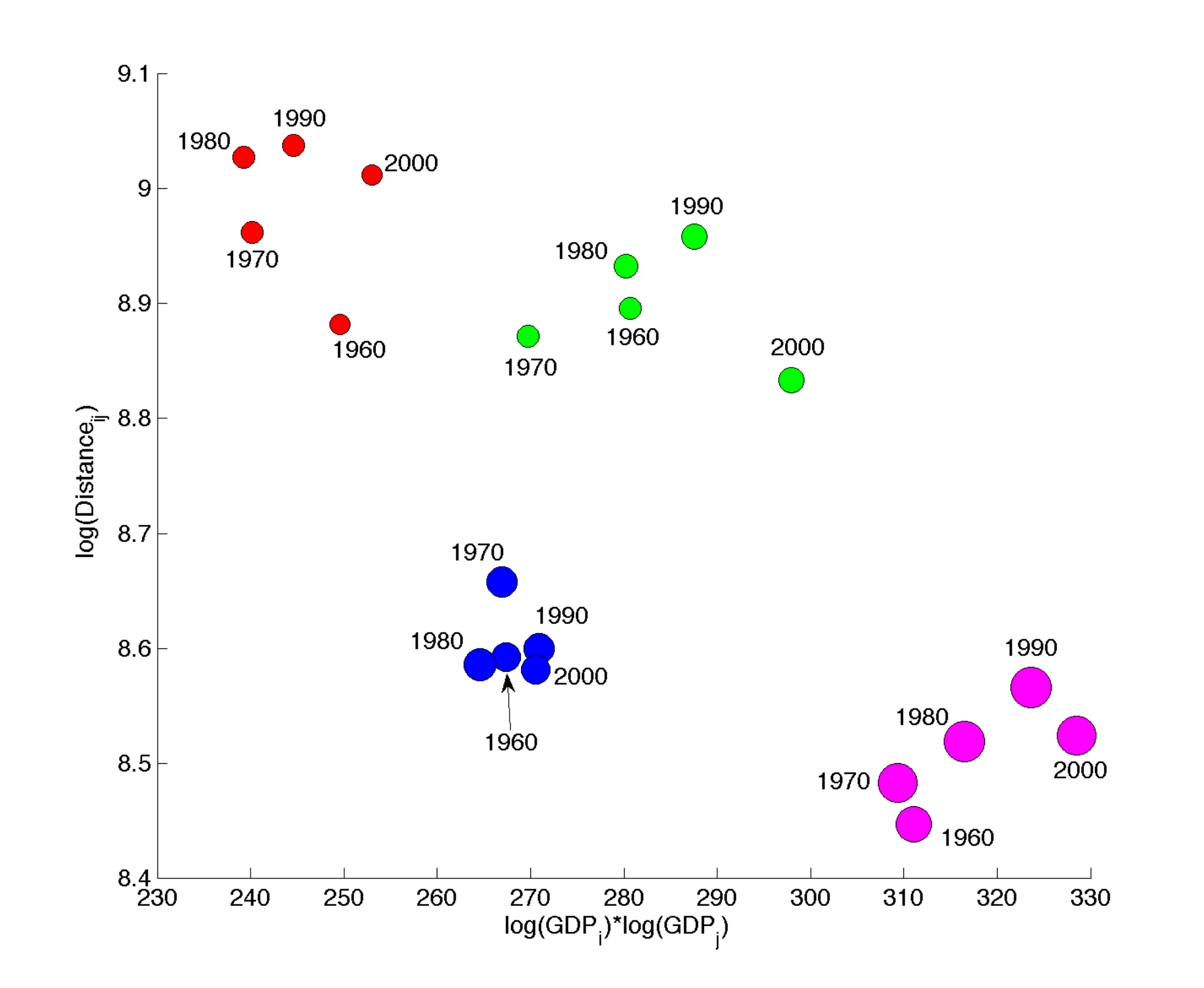}
\caption{Scatter plot of average $\log$(rGDP$_i^y)\cdot \log$(rGDP$_j^y$) versus average $\log$($\delta_{ij}$) conditional on matches/mismatches between IMN vs ITN binary structures. Colors: Red=Absence of link in both ITN and IMN.  Green=No link in IMN, link in ITN. Blue=No link in ITN, link in IMN. Magenta=Link in both ITN and IMN. Marker size is proportional to the product of standard deviations of $\log$(GDP$_i^y)*\log$(GDP$_j^y$) and $\log$($\delta_{ij}$), conditional to matches/mismatches between IMN vs ITN binary structures.}
\label{colored_balls}
\end{figure}

To see if real GDP and distances can also explain matches and mismatches between binary structures, we plot for each year the averages of the quantities $q_{ij}^y=\log$(rGDP$_i^y)\cdot \log$(rGDP$_j^y$) and $\log$($\delta_{ij}$), conditional to the four possible cases (depicted with different colors), namely: (i) no link in both IMN and ITN (red); (ii) link in ITN and no link in the IMN (green); (iii) link in IMN and no link in the ITN (blue); (iv) link in both ITN and IMN (magenta), see Figure \ref{colored_balls}. It is easy to see that a simultaneous absence vs presence of a link is due to the combination of, respectively, low rGDPs and high distances vs high rGDPs and short distances. Furthermore, as expected, the IMN is more sensible to distance than the ITN: a link in the ITN that is not present in the IMN is typically associated to large distances. On the contrary, the ITN is more sensible to rGDP. Even at smaller distances, country size plays a difference: when the latter is small enough, links in the IMN tend not to appear in the ITN. Note also that these results are very robust across time (all same-color dots are very close to each other) and display quite a good precision (cf. the relatively small conditional dispersion, i.e. colored balls do not overlap). Similar findings are obtained when rGDP is replaced by country population.  

\begin{figure}[h]
\hskip -0.5cm
\subfloat[]{\label{fig:instrength}\includegraphics[width=1.7in]{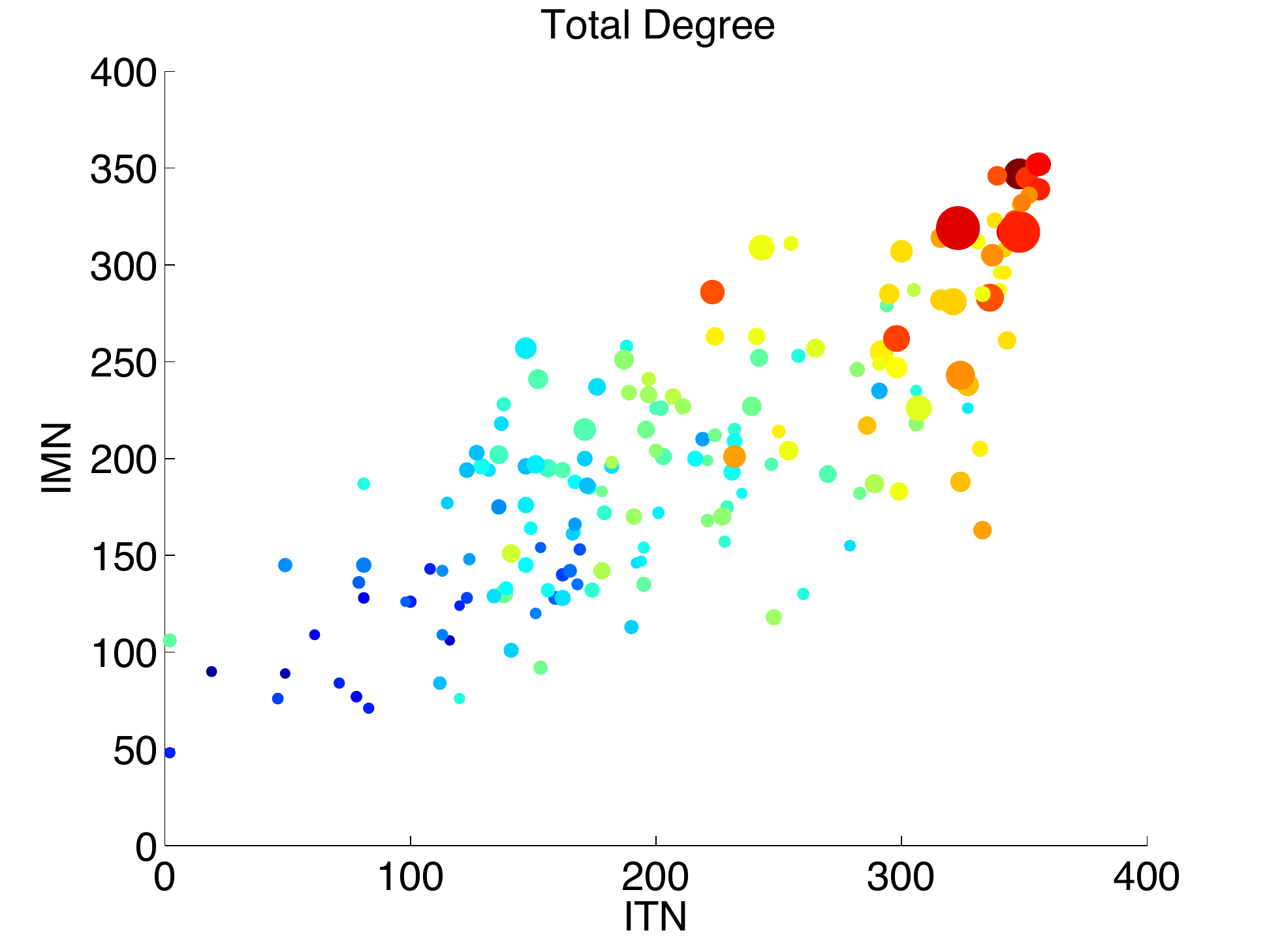}} 
\subfloat[]{\label{fig:outstrength}\includegraphics[width=1.7in]{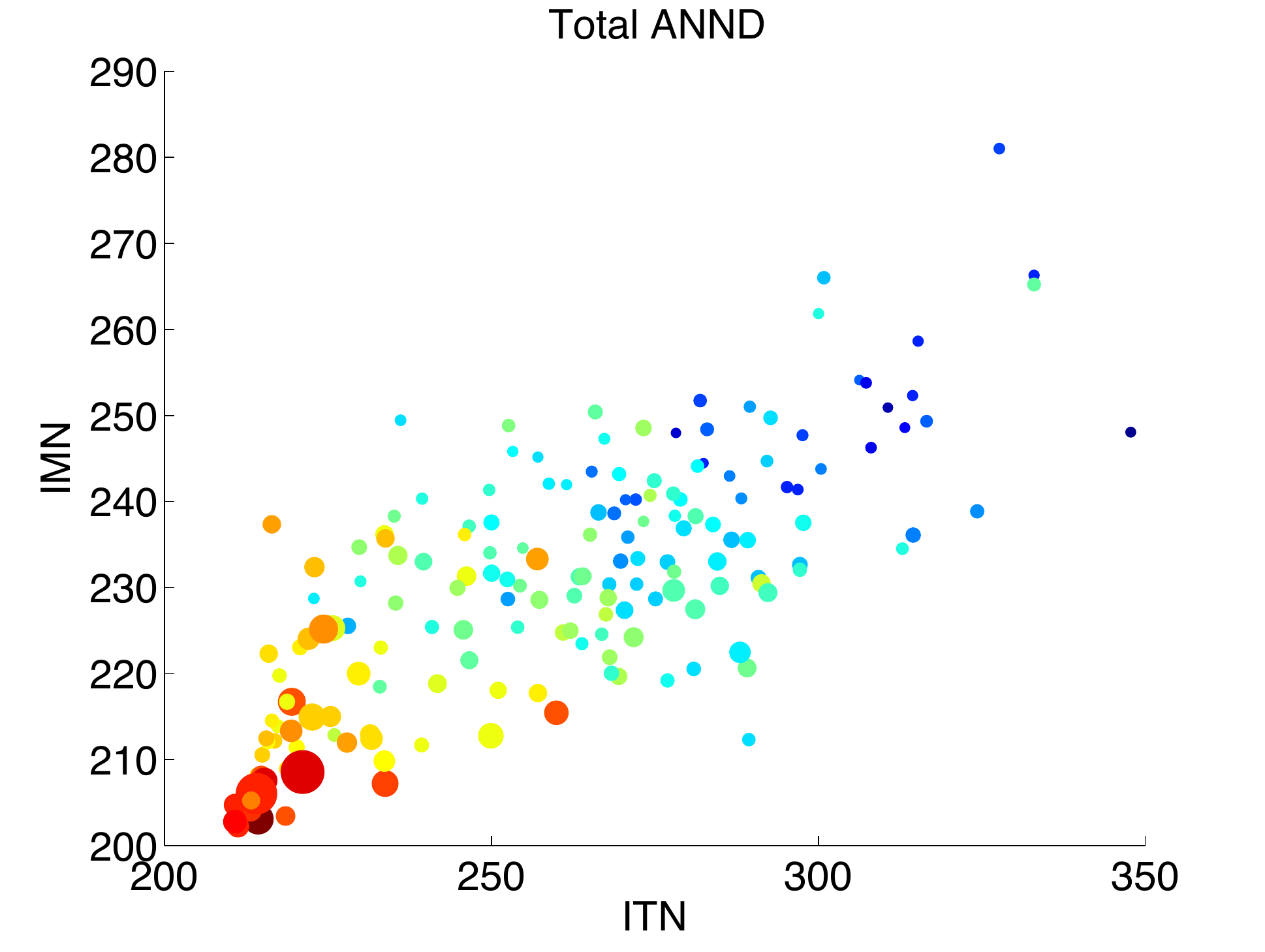}}\\
\par \hskip -0.5cm
\subfloat[]{\label{fig:instrength}\includegraphics[width=1.7in]{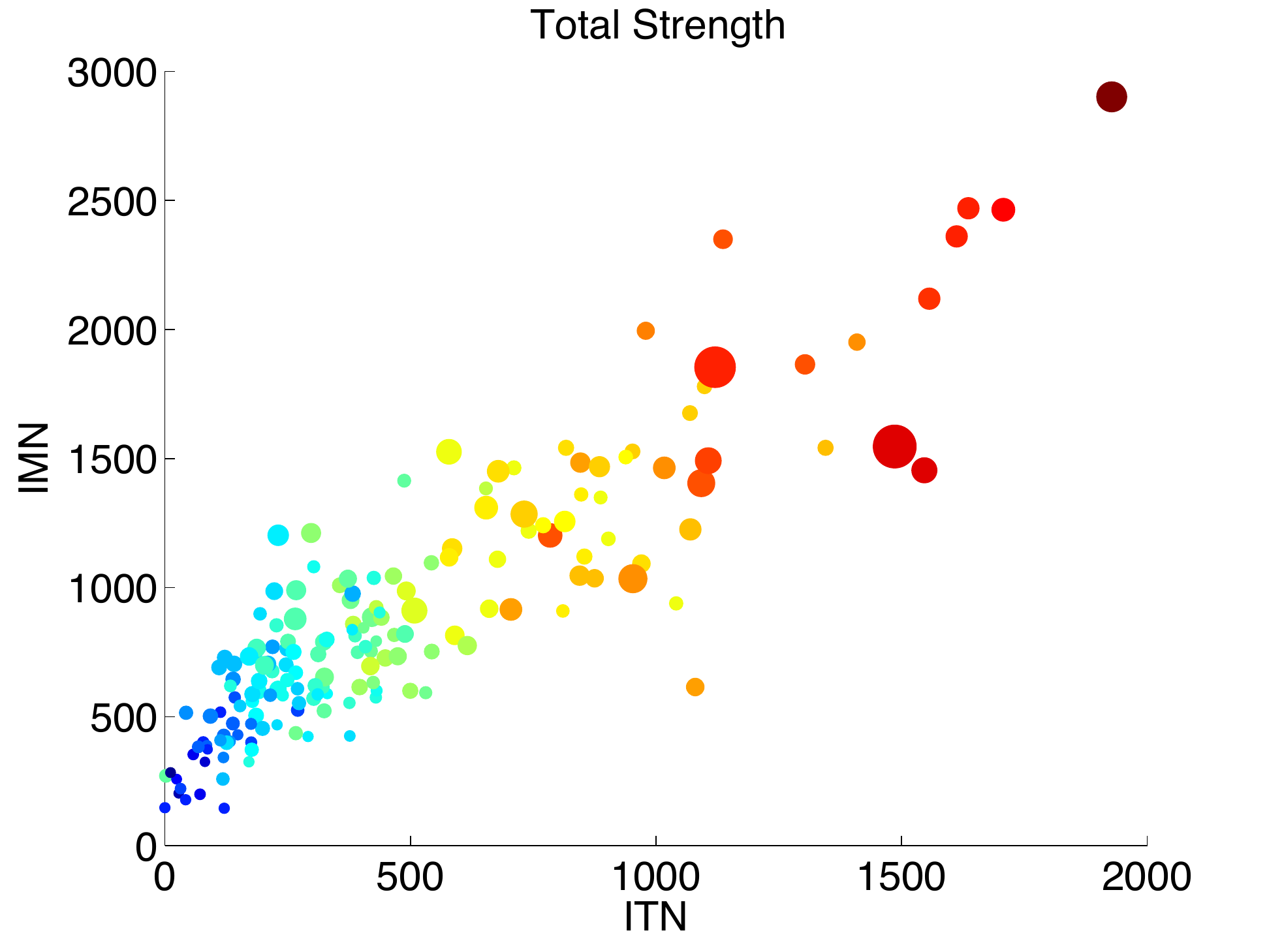}}
\subfloat[]{\label{fig:outstrength}\includegraphics[width=1.7in]{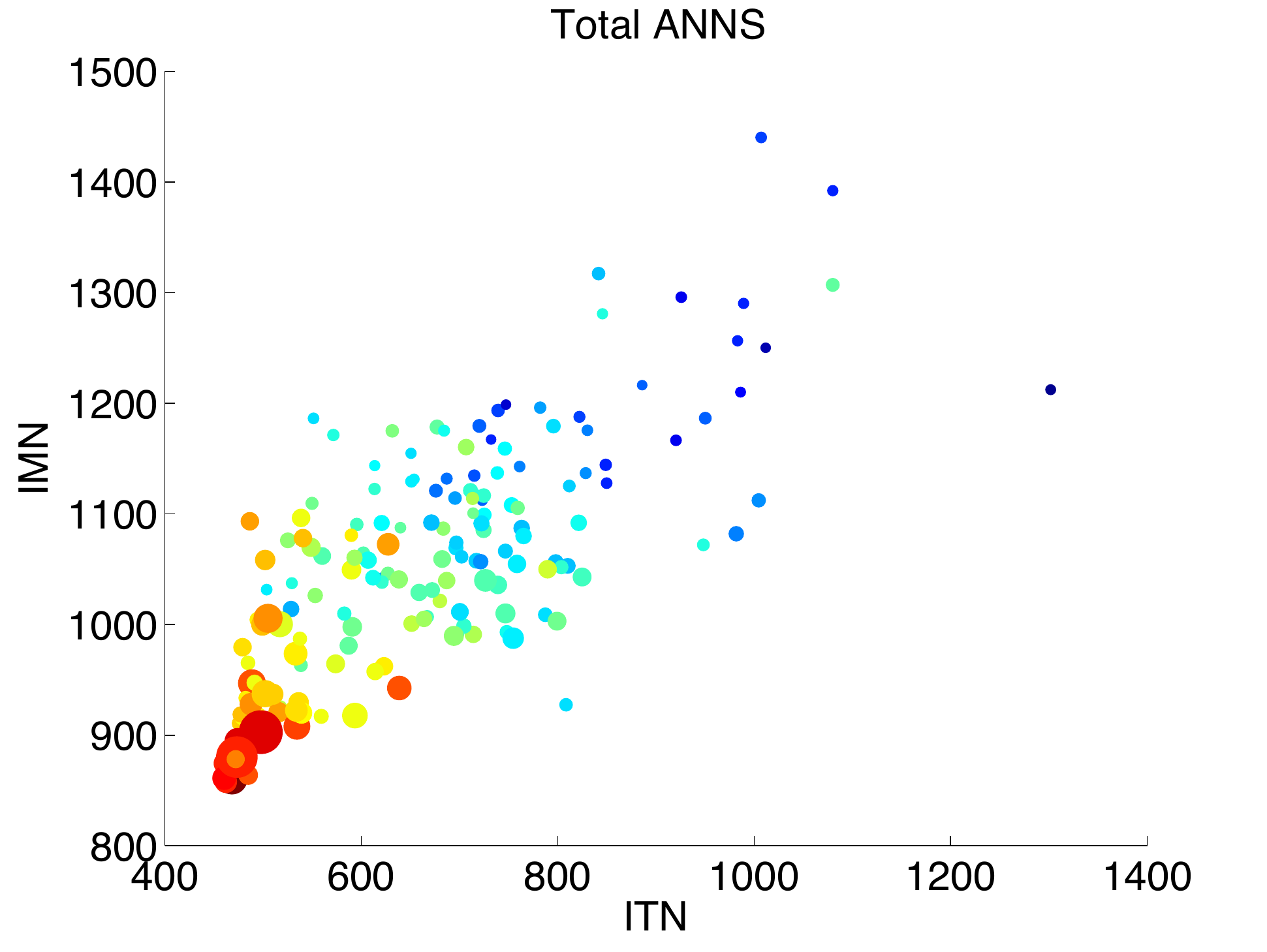}}
\caption{Correlation of node network statistics between IMN and ITN in year 2000. (a) Total degree; (b) Average nearest-neighbor degree (ANND). (c) Total strength. (d) Average nearest-neighbor strength (ANNS). Marker size is proportional to logs of POP$_{i}^y$. Colors scale (blue to red) is from lower to higher values logged GDP$_{i}^y$.}
\label{nodestats}
\end{figure}

Next, we explore whether node (binary and weighted) network statistics are correlated between the two layers of the IMTN. The four panels of Figure \ref{nodestats} summarize some of our findings for year 2000. Here we focus on four node network statistics\footnote{Similar findings hold for the whole range of network statistics that we have computed, including binary and weighted clustering \cite{Saramaki2006,Fa07}, hub and authority centrality \cite{Kleinberg_1999}, etc. The complete set of results are available from the Authors upon request.}: (i) total degree: the sum of inward and outward links of a node; (ii) total strength: sum of inward and outward link weights of a node; (iii)-(iv) total average nearest-neighbor degree (ANND) and strength (ANNS): average of node degree (respectively, strength) of the neighbors of a node, no matter the directionality of the links held by the node. Whereas total degree and ANND are computed on the binary IMTN, node strength and ANNS employ its weighted representation, where as customary in this literature \cite{Fa08,FagMas13}, logs of link-weights are used instead of levels to reduce the range of variation. As we did above, we correlate network statistics with country population (size of dots) and rGDP (color of dots, from blue to red). 

We find that both node degrees and strengths are positively and linearly related in the two layers, see panels (a) and (c). This means that if a country has more trade channels (respectively, trades more), it also carries more migration channels (respectively, holds larger immigrant/emigrant stocks). Again, it is easy to see that this positive relation is mostly explained by country demographic and economic size. We also find that if a country trades with countries that trade with many other partners or trade a lot, is also connected to countries that hold a lot of migration channels or stocks, i.e.  both ANND and ANNS are positively correlated in the two layers, cf. panels (b) and (d). 

\begin{figure}[h]
\hskip -0.5cm
\subfloat[]{\includegraphics[width=1.7in]{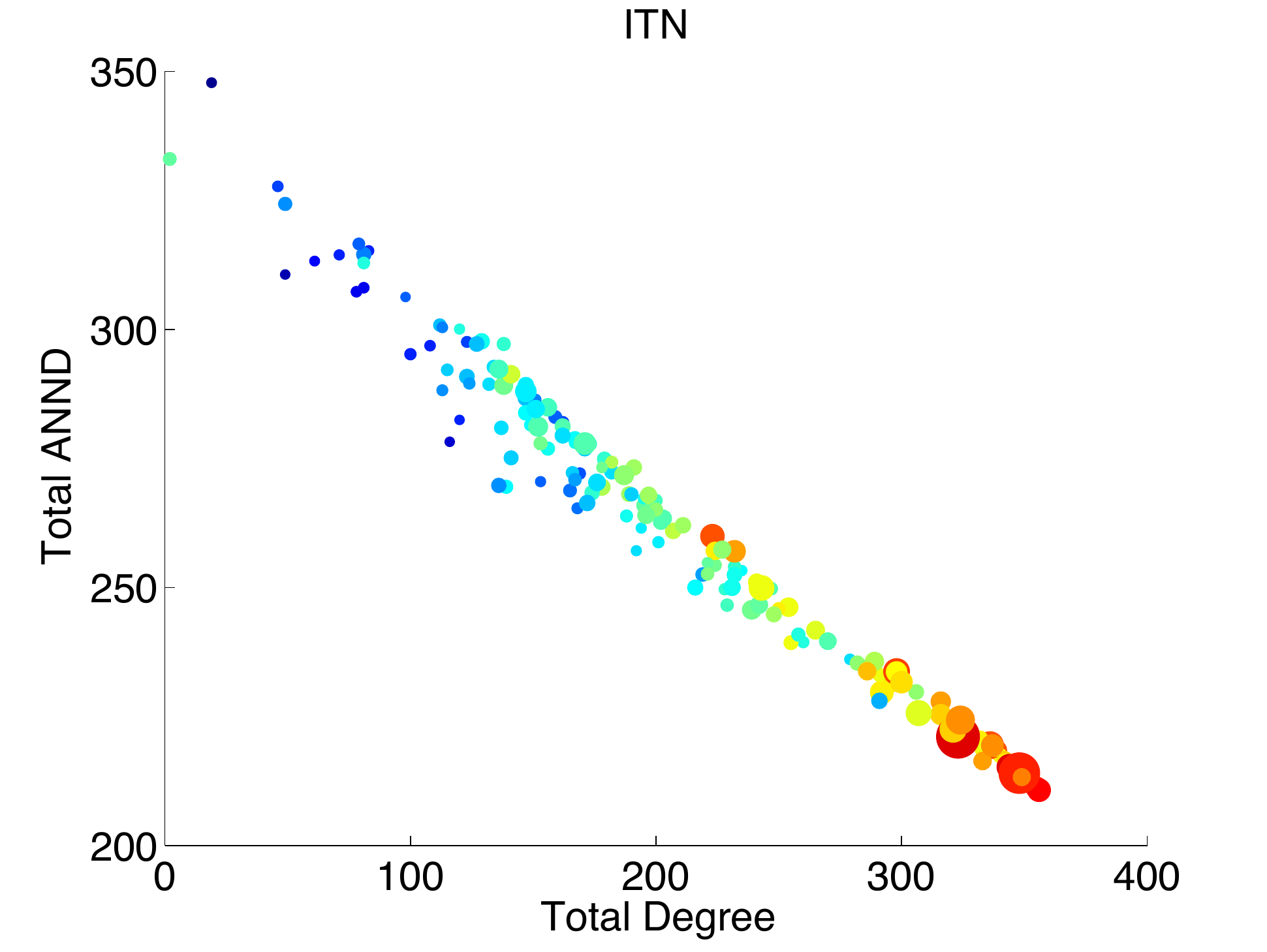}} 
\subfloat[]{\includegraphics[width=1.7in]{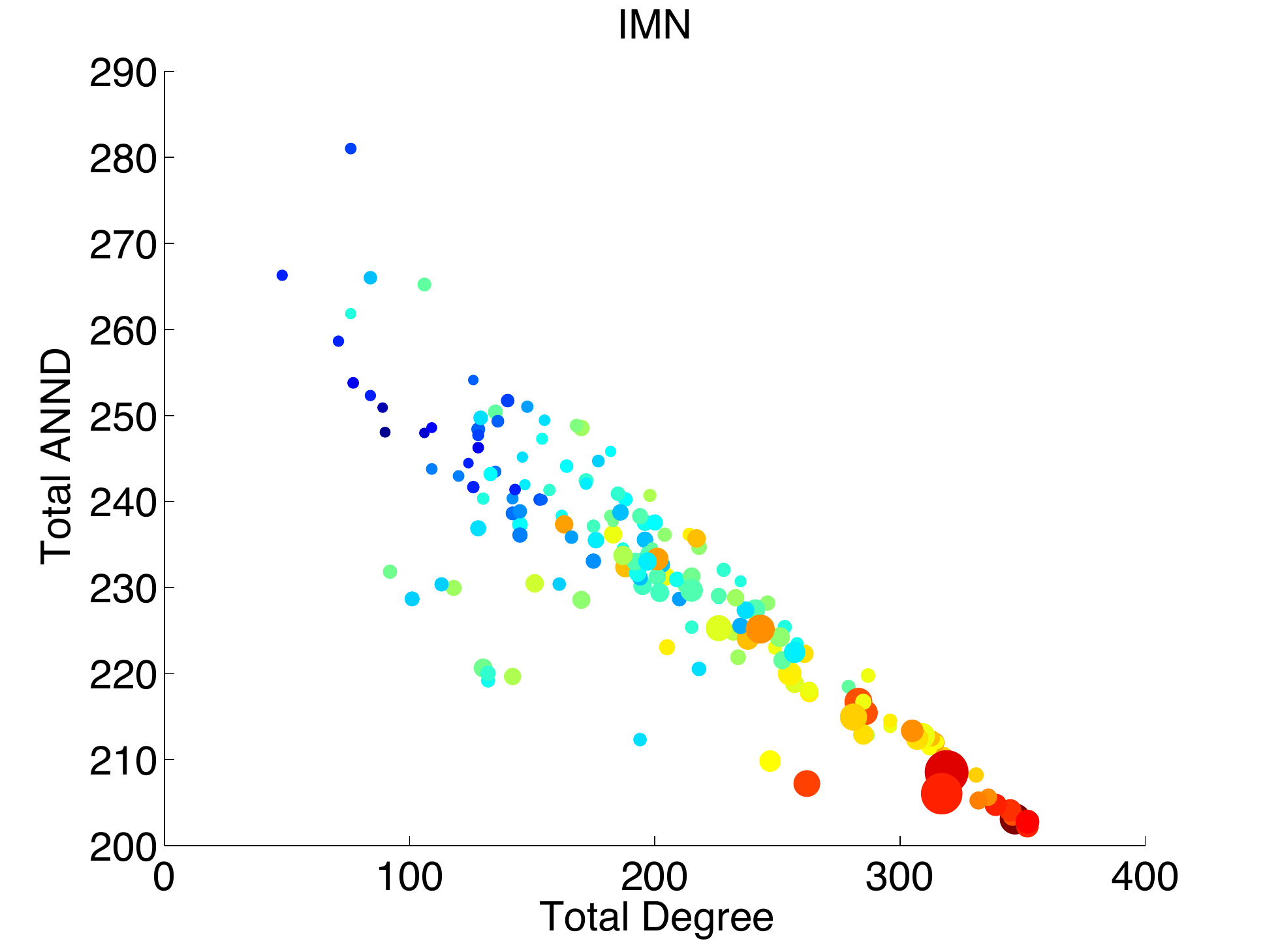}}\\
\par \hskip -0.5cm
\subfloat[]{\includegraphics[width=1.7in]{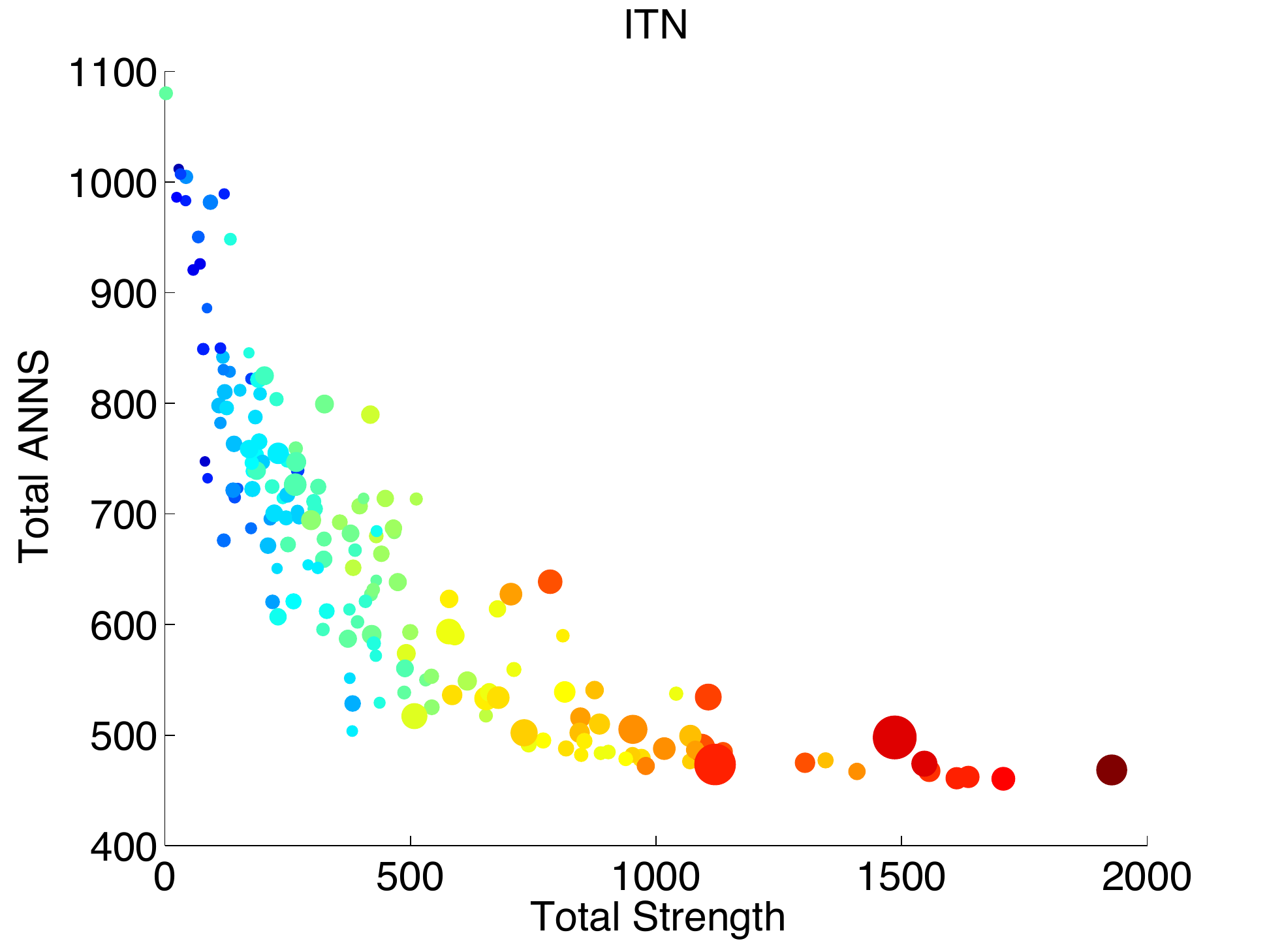}}
\subfloat[]{\includegraphics[width=1.7in]{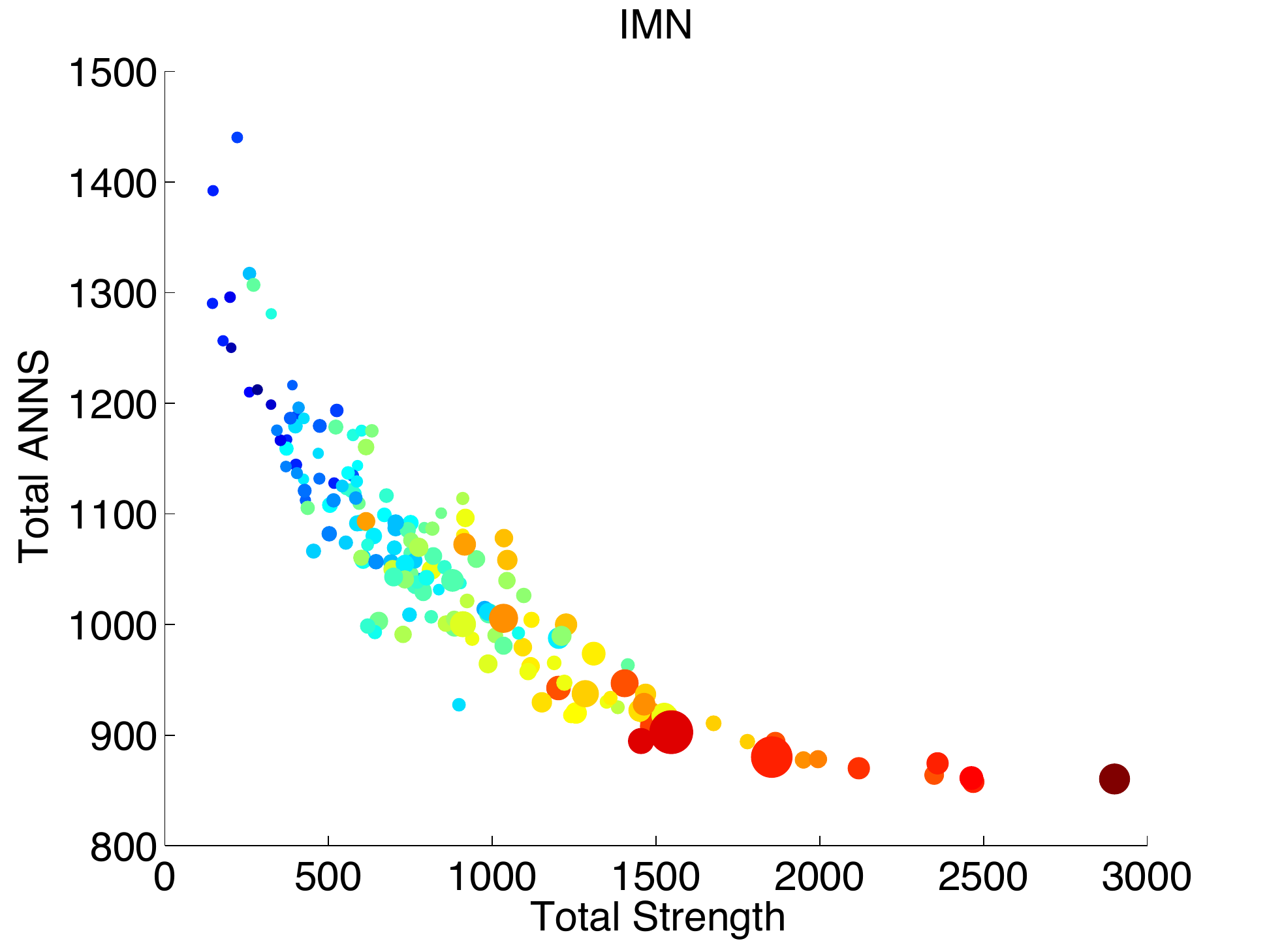}}
\caption{Disassortativity patterns within IMN and ITN in year 2000. Marker size is proportional to logs of POP$_{i}^y$. Colors scale (blue to red) is from lower to higher values logged GDP$_{i}^y$.}
\label{disassortativity}
\end{figure}

However, unlike what happens for degrees and strength, smaller levels of ANND and ANNS in the IMTN are associated to larger demographic and economic country sizes. To see why this is the case, we study binary and weighted disassortativity patterns \textit{within} the two IMTN layers. Figure \ref{disassortativity} scatter-plots node total degree (respectively, strength) vs ANND (respectively, ANNS), separately for ITN and IMN, and correlates this information with country population and rGDP as in Figure \ref{nodestats}. As already known, both networks display a marked (binary and weighted) disassortative behavior: the partners of more strongly connected nodes are weakly connected. However, larger countries (i.e. with higher levels of rGDP and POP) also hold larger degrees and strengths. Therefore, countries with larger levels of ANND and ANNS are smaller, in both economic and demographic terms. These results consistently hold also for the other years in the sample.  

\section{Does Migration Affect Trade? Regression Analyses using Network Statistics\label{regressions}}

In the preceding Section, we have explored the patterns of correlation between the two layers of the ITMN and their determinants. We move now to assessing whether there exist any causal relationship between the IMN and the ITN. As we have already noted discussing the evidence on binary structures (see Fig. \ref{bin_percentages}), the emergence of links in the ITN seems to be driven by existing migration corridors. More generally, many empirical studies find quite a robust evidence suggesting that bilateral migration affects international-trade flows \cite{GastonNelson_2011,Egger_etal_2012}. As argued by \cite{Gould_1994}, trade between any two countries $(i,j)$ may be enhanced by the presence of stock of immigrants present in either country and coming from the other one ($m_{ji}$ and $m_{ij}$). This is because migrants originating in $j$ and present in $i$ (and vice versa) may foster imports of goods produced in their mother country (bilateral consumption-preference effect) or reduce import transaction costs thanks to their better knowledge of both home- and host-country laws, habits, and regulations (bilateral information effect). 

However, one may posit that trade between any two countries can be fostered not only by bilateral-migration effects, but also thanks to migrants coming from other ``third parties'' and, more generally, by the overall connectivity and centrality of both countries in the IMN. This is because the better a pair of countries is connected in the IMN, the larger the average number of third countries that they share as origin of immigration flows and the more likely the presence of strong third-party migrant communities in both countries. This may further enhance trade via both preference and information effects. Moreover, it may happen that two countries are relatively well connected in the IMN (in both binary and weighted terms) even if they share a very limited number of non-overlapping third parties. In such a case, one may ask whether a cosmopolitan environment engendered by the presence of many ethnic groups in both countries can be trade enhancing ---and if so why.

In order to explore these issues, we perform a set of econometric exercises using a standard gravity-model of trade, expanded to take into account migration network effects. Building on Ref. \cite{Parsons_2012}, we fit a gravity model whose general specification reads: 
\begin{equation}
\log \tau_{ij}^y=\kappa + \phi_i^y+\gamma_j^y+\alpha \log(\delta_{ij})+ \boldsymbol{\beta} \boldsymbol{Z}_{ij}^y+ \boldsymbol{\mu} \boldsymbol{W}_{ij}^y+\varepsilon_{ij}^y
\end{equation}
where $\varepsilon_{ij}^y$ is the error term; $\kappa$ is a constant; $\tau_{ij}^y=t^y_{ij}+t^y_{ji}$ is total bilateral trade; $(\phi_i^y,\gamma_j^y)$ are country-time importer-exporter dummies controlling for all country-specific variables such as rGDP and POP; $\delta_{ij}$ is geographical distance; $\boldsymbol{Z}_{ij}^y$ features bilateral country dummies (CONTIG, COMLANG, RTA$^y$)\footnote{Results are robust to additional controls such as common religion, common colonial ties, and landlocking effects.}; and $\boldsymbol{W}_{ij}^y$ is a vector of migration-related network variables accounting for bilateral and common vs. non-overlapping third-country effects. 

We test five different econometric specifications. In the first two, we only control for baseline gravity-related variables and total bilateral migration stock ---i.e. BIL\_MIG$_{ij}^y$=$\log(m_{ij}^y+m_{ji}^y)$ --- whereas in the remaining specifications we add also network effects related to country centrality in the IMN. In our exercises, we employ as a measure of country centrality \textit{in-degree country centralization}, defined as:
\begin{equation}
\text{IN\_CENTR}_{i}^y=\frac{ind_i^{y}}{N^y}
\end{equation} 
where $ind_i^{y}$ is country in-degree (i.e. the number of inward links of country $i$). This is because we expect inward migration to be relevant in explaining bilateral trade rather than outward channels. Notice also that in-degree centralization is highly and positively correlated with all other (binary and weighted) centrality indicators in the IMN (i.e. eigenvector-based indicators, betweenness centrality, etc.). For this reason, our results are quite robust to alternative centrality measures. We choose to employ binary centralization indicators instead of weighted-centrality measures to reduce endogeneity issues. Indeed, in principle, the error term may be correlated with the explanatory variables due to a reverse-causation link going from trade to migration. We argue that this problem may be almost irrelevant in terms of migration channels, as it is very unlikely that changes in bilateral-trade levels may destroy or form new links in the IMN.\footnote{Endogeneity issues may also bias estimation of the coefficients of total bilateral migration stocks. Notice, however, that whereas the dependent variable refers to total trade in year $y$, total bilateral migration records migrants present in year $y$ in the host country, but refers to flows occurred in the preceeding years, thus weakining the reverse-causation link from trade to migration in our exercises. More on this in Section \ref{conclusions}.} Since we employ importer-exporter time dummies, we add the log of the sum of country $i$ and $j$ in-degree centralization:
\begin{equation}
\text{IN\_CENTR}_{ij}^y=\text{IN\_CENTR}_{i}^y+\text{IN\_CENTR}_{j}^y,
\end{equation} 
instead of the two separately. Furthermore, in order to separate third-party common and non-overlapping inward migration channels we also use in our regressions the log of the share of common in-neighbors of any given pairs of countries (COMM\_IN$_{ij}^y$) and the log of the share of inward channels that the two countries do not share (NONCOMM\_IN$_{ij}^y$), where shares are computed dividing by $N^y$.

\begin{table*}[ht]\centering
\renewcommand{\arraystretch}{1.3}
\label{table_gravity}
\begin{tabular}{l*{5}{c}}
\hline\hline
            &\multicolumn{1}{c}{(1)}&\multicolumn{1}{c}{(2)}&\multicolumn{1}{c}{(3)}&\multicolumn{1}{c}{(4)}  &\multicolumn{1}{c}{(5)}\\
\hline
$\log$($\delta_{ij}$)   &      -1.146***&      -0.745***&      -0.740***&      -0.740***&      -0.740***\\
CONTIG      &       0.511***&       0.132***&       0.141***&       0.139***&       0.139***\\
COMLANG &       0.529***&       0.320***&       0.319***&       0.316***&       0.316***\\
RTA$^y$          &       0.434***&       0.197***&       0.178***&       0.187***&       0.187***\\
$\log$(BIL\_MIG$_{ij}^y$) &               &       0.109***&       0.102***&       0.113***&       0.104***\\
$\log$(IN\_CENTR$_{ij}^y$)  &               &               &       0.994***&               &               \\
$\log$(COMM\_IN$_{ij}^y$) &               &               &               &       0.211***&       0.343***   \\
$\log$(NONCOMM\_IN$_{ij}^y$) &               &               &               &               &       0.066***   \\
\hline
\(N\)       &       58812   &       34404   &       34404   &       34398   &       34398   \\
\(R^{2}\)   &       0.749   &       0.798   &       0.812   &       0.807   &       0.842   \\
\hline\hline
\end{tabular} 
\caption{Gravity-model estimation. Full-sample (pooled) ordinary least-square (OLS) fit. Years $y=1960,\dots,2000$. Dependent variable: logs of total bilateral trade $\tau_{ij}^y=t^y_{ij}+t^y_{ji}$. Country-year dummy variables for importer/exporter effects and constant included. $N$=No. of Observations. Explanatory variables: See main text. Significance levels: $^{***}=1\%, ^{**}=5\%, ^{*}=10\%$.\label{table_gravity}}
\end{table*}

In Table \ref{table_gravity} we show the results of OLS estimations. All specifications attain a very high goodness of fit as it always happens in empirical gravity estimation. The addition of network statistics induces an increase in the $R^2$, albeit limited. The impact of distance, contiguity, common language and participation to a trade agreement are strong, significant, and signed in line with existing studies. Total bilateral migration positively affects bilateral trade as expected, and its impact is almost constant no matter the chosen specification \cite{Parsons_2012}. Columns (3)-(5) report regressions where country-network centrality indicators are accounted for. We find that the more total inward-migration corridors a pair of country holds, the larger their bilateral trade, i.e. IN\_CENTR$_{ij}^y$ has a positive and significant effect on trade. To check whether this is due to common vs non-overlapping in-neighboring channels, columns (4) and (5) report specifications where only COMM\_IN$_{ij}^y$ or both COMM\_IN$_{ij}^y$ and NONCOMM\_IN$_{ij}^y$ enter the model. Estimates suggest that: (i) common third parties have a positive effect on bilateral trade; (ii) once one controls for common third-parties, non-overlapping channels are also trade enhancing, even if with a smaller impact\footnote{These results hold true also if: (i) country-time importer-exporter dummies are removed and replaced with country rGDPs; (ii) we employ $t^y_{ij}$ as dependent variable and we separately add as regressors country-centrality indicators ($\text{IN\_CENTR}_{i}^y$ and $\text{IN\_CENTR}_{j}^y$). Note also that the positive effect on trade of NONCOMM\_IN$_{ij}^y$ is preserved when one enters this variable in the regression without COMM\_IN$_{ij}^y$.}.

The foregoing evidence suggests that in addition to bilateral-migration effects, trade between any two countries $(i,j)$ may increase due to their connectivity in the binary IMN. This might happen via two related mechanisms. First, pairs of countries holding more inward links are more likely to share an increasing number of common third-party migration origins $k\neq (i,j)$. Everything else being equal, this implies a larger number of migrants coming from the same origin country $k$ that are commonly shared by $(i,j)$ and, therefore, thanks to consumption-preference and information effects, more bilateral trade \cite{Rau99,Fel10,Fel12}. Second, a smaller but still significant trade-enhancing effect can come from the presence in both countries of an increasing number of inward migration corridors that are however not shared by $i$ and $j$. In other words, if countries $i$ and $j$ host migrants originated respectively from countries $I=\{i_1,\dots,i_m\}$ and $J=\{j_1,\dots,j_n\}$, with $I\bigcap J=\O$, the larger $m$ and/or $n$, the higher bilateral trade between the two countries. This second trade-enhancing effect can have a twofold explanation. On the one hand, more non-overlapping migration channels, coupled with commonly-shared origins, may imply more cosmopolitan and inclusive environments in both countries, which may in turn foster, in all ethnic groups, learning processes about consumption patterns of ethnic groups commonly shared by the two countries, and therefore more bilateral trade. On the other hand, more non-overlapping inward migration channels imply a higher probability to find in both countries more second-generation migrants belonging to the same ethnic group. Indeed, our data record migrants according their birth-place and not necessarily their ethnic origin. Therefore, it may be the case that, even if countries $h_i$ and $h_j$ are not shared as inward channels by $i$ and $j$ respectively, they can send second-generation migrants belonging to the same ethnic group to $i$ and $j$, thus enhancing their bilateral trade. This effect cannot be entirely picked up by COMM\_IN$_{ij}^y$ and it can thus show up, as Table \ref{table_gravity} suggests, in the coefficient of NONCOMM\_IN$_{ij}^y$.

\section{Conclusions\label{conclusions}}
This paper has explored the relationships between international migration and trade using a complex-network approach. More specifically, we have performed two related exercises. First, we have investigated the patterns of correlation between the ITN and the IMN, comparing link weights, topological structures and node network statistics. We have found that trade and migration networks are strongly correlated and such relation can be mostly explained by country economic and demographic size and geographical distance. Second, we have asked whether country centrality in the IMN can explain bilateral trade. Expanding upon the existing literature in economics, we have fit to the data gravity models of bilateral trade adding migration-network variables among the regressors. These control for country in-degree centralization, and the number of common vs. non-overlapping inward migration channels. Our results indicate that the larger the number of inward ---both common and non-overlapping--- migration corridors held by any two countries, the higher bilateral trade.

Our work can be extended in at least two directions. First, one can further investigate endogeneity issues arising in gravity-model exercises, due to the reverse-causation link possibly existing from trade to migration. A possible way out might involve instrumenting migration stocks (e.g., using a simple migration gravity model) and replace migration-related regressors with the predictions from the instrumental-variable estimation. Second, one can build upon the idea of a macroeconomic multi-network and add to the picture additional network layers related e.g. to temporary international human-mobility and foreign-direct investment. This may allow one to better understand the importance of international migration and human-mobility patterns in explaining macroeconomic variables. 

\section*{Acknowledgment}
\noindent G.F. gratefully acknowledges support received by the research project “The International Trade Network: Empirical Analyses and Theoretical Models” funded by the Italian Ministry of Education, University and Research (Scientific Research Programs of National Relevance 2009).


\begin{thebibliography}{38}
\expandafter\ifx\csname natexlab\endcsname\relax\def\natexlab#1{#1}\fi
\expandafter\ifx\csname bibnamefont\endcsname\relax
  \def\bibnamefont#1{#1}\fi
\expandafter\ifx\csname bibfnamefont\endcsname\relax
  \def\bibfnamefont#1{#1}\fi
\expandafter\ifx\csname citenamefont\endcsname\relax
  \def\citenamefont#1{#1}\fi
\expandafter\ifx\csname url\endcsname\relax
  \def\url#1{\texttt{#1}}\fi
\expandafter\ifx\csname urlprefix\endcsname\relax\def\urlprefix{URL }\fi
\providecommand{\bibinfo}[2]{#2}
\providecommand{\eprint}[2][]{\url{#2}}

\bibitem[{\citenamefont{Schweitzer et~al.}(2009)\citenamefont{Schweitzer,
  Fagiolo, Sornette, Vega-Redondo, Vespignani, and White}}]{ScienceNets2009}
\bibinfo{author}{\bibfnamefont{F.}~\bibnamefont{Schweitzer}},
  \bibinfo{author}{\bibfnamefont{G.}~\bibnamefont{Fagiolo}},
  \bibinfo{author}{\bibfnamefont{D.}~\bibnamefont{Sornette}},
  \bibinfo{author}{\bibfnamefont{F.}~\bibnamefont{Vega-Redondo}},
  \bibinfo{author}{\bibfnamefont{A.}~\bibnamefont{Vespignani}},
  \bibnamefont{and} \bibinfo{author}{\bibfnamefont{D.~R.} \bibnamefont{White}},
  \bibinfo{journal}{Science} \textbf{\bibinfo{volume}{325}},
  \bibinfo{pages}{422} (\bibinfo{year}{2009}).

\bibitem[{\citenamefont{Serrano and Bogu\~n\'a}(2003)}]{Se03}
\bibinfo{author}{\bibfnamefont{A.}~\bibnamefont{Serrano}} \bibnamefont{and}
  \bibinfo{author}{\bibfnamefont{M.}~\bibnamefont{Bogu\~n\'a}},
  \bibinfo{journal}{Physical Review E} \textbf{\bibinfo{volume}{68}}
  (\bibinfo{year}{2003}).

\bibitem[{\citenamefont{Garlaschelli and Loffredo}(2004)}]{Ga04}
\bibinfo{author}{\bibfnamefont{D.}~\bibnamefont{Garlaschelli}}
  \bibnamefont{and} \bibinfo{author}{\bibfnamefont{M.}~\bibnamefont{Loffredo}},
  \bibinfo{journal}{Physical Review Letters.} \textbf{\bibinfo{volume}{93}}
  (\bibinfo{year}{2004}).

\bibitem[{\citenamefont{Fagiolo et~al.}(2008)\citenamefont{Fagiolo, Schiavo,
  and Reyes}}]{Fa08}
\bibinfo{author}{\bibfnamefont{G.}~\bibnamefont{Fagiolo}},
  \bibinfo{author}{\bibfnamefont{S.}~\bibnamefont{Schiavo}}, \bibnamefont{and}
  \bibinfo{author}{\bibfnamefont{J.}~\bibnamefont{Reyes}},
  \bibinfo{journal}{Physica A} \textbf{\bibinfo{volume}{387}},
  \bibinfo{pages}{3868–73} (\bibinfo{year}{2008}).

\bibitem[{\citenamefont{Fagiolo et~al.}(2009)\citenamefont{Fagiolo, Schiavo,
  and Reyes}}]{Fa09}
\bibinfo{author}{\bibfnamefont{G.}~\bibnamefont{Fagiolo}},
  \bibinfo{author}{\bibfnamefont{S.}~\bibnamefont{Schiavo}}, \bibnamefont{and}
  \bibinfo{author}{\bibfnamefont{J.}~\bibnamefont{Reyes}},
  \bibinfo{journal}{Physical Review E} \textbf{\bibinfo{volume}{79}}
  (\bibinfo{year}{2009}).

\bibitem[{\citenamefont{Fagiolo et~al.}(2010)\citenamefont{Fagiolo, Schiavo,
  and Reyes}}]{Fa10}
\bibinfo{author}{\bibfnamefont{G.}~\bibnamefont{Fagiolo}},
  \bibinfo{author}{\bibfnamefont{S.}~\bibnamefont{Schiavo}}, \bibnamefont{and}
  \bibinfo{author}{\bibfnamefont{J.}~\bibnamefont{Reyes}},
  \bibinfo{journal}{Journal of Evolutionary Economics}
  \textbf{\bibinfo{volume}{20}}, \bibinfo{pages}{479} (\bibinfo{year}{2010}).

\bibitem[{\citenamefont{Song et~al.}(2009)\citenamefont{Song, Jiang, and
  Zhou}}]{Song2009}
\bibinfo{author}{\bibfnamefont{D.~M.} \bibnamefont{Song}},
  \bibinfo{author}{\bibfnamefont{Z.~Q.} \bibnamefont{Jiang}}, \bibnamefont{and}
  \bibinfo{author}{\bibfnamefont{W.~X.} \bibnamefont{Zhou}},
  \bibinfo{journal}{Physica A} \textbf{\bibinfo{volume}{388}},
  \bibinfo{pages}{2450} (\bibinfo{year}{2009}).

\bibitem[{\citenamefont{Schiavo et~al.}(2010)\citenamefont{Schiavo, Reyes, and
  Fagiolo}}]{finnet_qf_2010}
\bibinfo{author}{\bibfnamefont{S.}~\bibnamefont{Schiavo}},
  \bibinfo{author}{\bibfnamefont{J.}~\bibnamefont{Reyes}}, \bibnamefont{and}
  \bibinfo{author}{\bibfnamefont{G.}~\bibnamefont{Fagiolo}},
  \bibinfo{journal}{Quantitative Finance} \textbf{\bibinfo{volume}{10}},
  \bibinfo{pages}{389} (\bibinfo{year}{2010}).

\bibitem[{\citenamefont{Minoiu and Reyes}(2013)}]{minoiu_reyes_2013}
\bibinfo{author}{\bibfnamefont{C.}~\bibnamefont{Minoiu}} \bibnamefont{and}
  \bibinfo{author}{\bibfnamefont{J.~A.} \bibnamefont{Reyes}},
  \bibinfo{journal}{Journal of Financial Stability}
  \textbf{\bibinfo{volume}{9}}, \bibinfo{pages}{168} (\bibinfo{year}{2013}).

\bibitem[{\citenamefont{Chinazzi and Fagiolo}(2013)}]{contagion_survey}
\bibinfo{author}{\bibfnamefont{M.}~\bibnamefont{Chinazzi}} \bibnamefont{and}
  \bibinfo{author}{\bibfnamefont{G.}~\bibnamefont{Fagiolo}}, \bibinfo{type}{LEM
  Papers Series} \bibinfo{number}{2013/08}, \bibinfo{institution}{Laboratory of
  Economics and Management (LEM), Sant'Anna School of Advanced Studies, Pisa,
  Italy} (\bibinfo{year}{2013}),
  \urlprefix\url{http://ideas.repec.org/p/ssa/lemwps/2013-08.html}.

\bibitem[{\citenamefont{Fagiolo and Mastrorillo}(2013)}]{FagMas13}
\bibinfo{author}{\bibfnamefont{G.}~\bibnamefont{Fagiolo}} \bibnamefont{and}
  \bibinfo{author}{\bibfnamefont{M.}~\bibnamefont{Mastrorillo}},
  \bibinfo{journal}{Phisical Review E} \textbf{\bibinfo{volume}{88}},
  \bibinfo{pages}{012812} (\bibinfo{year}{2013}).

\bibitem[{\citenamefont{Davis et~al.}(2013)\citenamefont{Davis, D'Odorico,
  Laio, and Ridolfi}}]{Davis_etal_2013}
\bibinfo{author}{\bibfnamefont{K.~F.} \bibnamefont{Davis}},
  \bibinfo{author}{\bibfnamefont{P.}~\bibnamefont{D'Odorico}},
  \bibinfo{author}{\bibfnamefont{F.}~\bibnamefont{Laio}}, \bibnamefont{and}
  \bibinfo{author}{\bibfnamefont{L.}~\bibnamefont{Ridolfi}},
  \bibinfo{journal}{PLoS ONE} \textbf{\bibinfo{volume}{8}},
  \bibinfo{pages}{e53723} (\bibinfo{year}{2013}),
  \urlprefix\url{http://dx.doi.org/10.1371/journal.pone.0053723}.

\bibitem[{\citenamefont{Lee et~al.}(2011)\citenamefont{Lee, Yang, Kim, Lee,
  Goh, and Kim}}]{lee_etal_plosone_2010}
\bibinfo{author}{\bibfnamefont{K.}~\bibnamefont{Lee}},
  \bibinfo{author}{\bibfnamefont{J.}~\bibnamefont{Yang}},
  \bibinfo{author}{\bibfnamefont{G.}~\bibnamefont{Kim}},
  \bibinfo{author}{\bibfnamefont{J.}~\bibnamefont{Lee}},
  \bibinfo{author}{\bibfnamefont{K.}~\bibnamefont{Goh}}, \bibnamefont{and}
  \bibinfo{author}{\bibfnamefont{I.}~\bibnamefont{Kim}}, \bibinfo{journal}{PLoS
  One} \textbf{\bibinfo{volume}{6}}, \bibinfo{pages}{e18443}
  (\bibinfo{year}{2011}).

\bibitem[{\citenamefont{Chinazzi et~al.}(2013)\citenamefont{Chinazzi, Fagiolo,
  Reyes, and Schiavo}}]{postmortem_jedc}
\bibinfo{author}{\bibfnamefont{M.}~\bibnamefont{Chinazzi}},
  \bibinfo{author}{\bibfnamefont{G.}~\bibnamefont{Fagiolo}},
  \bibinfo{author}{\bibfnamefont{J.~A.} \bibnamefont{Reyes}}, \bibnamefont{and}
  \bibinfo{author}{\bibfnamefont{S.}~\bibnamefont{Schiavo}},
  \bibinfo{journal}{Journal of Economic Dynamics and Control}
  \textbf{\bibinfo{volume}{37}}, \bibinfo{pages}{1692} (\bibinfo{year}{2013}).

\bibitem[{\citenamefont{Gaston and Nelson}(2011)}]{GastonNelson_2011}
\bibinfo{author}{\bibfnamefont{N.}~\bibnamefont{Gaston}} \bibnamefont{and}
  \bibinfo{author}{\bibfnamefont{D.}~\bibnamefont{Nelson}}, in
  \emph{\bibinfo{booktitle}{Handbook of International Trade}}, edited by
  \bibinfo{editor}{\bibfnamefont{D.}~\bibnamefont{Bernhofen}},
  \bibinfo{editor}{\bibfnamefont{R.}~\bibnamefont{Falvey}},
  \bibinfo{editor}{\bibfnamefont{D.}~\bibnamefont{Greenaway}},
  \bibnamefont{and}
  \bibinfo{editor}{\bibfnamefont{U.}~\bibnamefont{Kreickemeier}}
  (\bibinfo{publisher}{Palgrave}, \bibinfo{address}{London},
  \bibinfo{year}{2011}).

\bibitem[{\citenamefont{Egger et~al.}(2012)\citenamefont{Egger, von Ehrlich,
  and Nelson}}]{Egger_etal_2012}
\bibinfo{author}{\bibfnamefont{P.~H.} \bibnamefont{Egger}},
  \bibinfo{author}{\bibfnamefont{M.}~\bibnamefont{von Ehrlich}},
  \bibnamefont{and} \bibinfo{author}{\bibfnamefont{D.~R.}
  \bibnamefont{Nelson}}, \bibinfo{journal}{The World Economy}
  \textbf{\bibinfo{volume}{35}}, \bibinfo{pages}{216} (\bibinfo{year}{2012}).

\bibitem[{\citenamefont{Parsons}(2012)}]{Parsons_2012}
\bibinfo{author}{\bibfnamefont{C.~R.} \bibnamefont{Parsons}},
  \bibinfo{type}{Policy Research Working Paper Series} \bibinfo{number}{6034},
  \bibinfo{institution}{The World Bank} (\bibinfo{year}{2012}).

\bibitem[{\citenamefont{Gould}(1994)}]{Gould_1994}
\bibinfo{author}{\bibfnamefont{D.~M.} \bibnamefont{Gould}},
  \bibinfo{journal}{The Review of Economics and Statistics}
  \textbf{\bibinfo{volume}{76}}, \bibinfo{pages}{302} (\bibinfo{year}{1994}).

\bibitem[{\citenamefont{Leloup}(1996)}]{Le96}
\bibinfo{author}{\bibfnamefont{F.}~\bibnamefont{Leloup}},
  \bibinfo{journal}{International Journal of Anthropology}
  \textbf{\bibinfo{volume}{11}}, \bibinfo{pages}{101} (\bibinfo{year}{1996}).

\bibitem[{\citenamefont{Garlaschelli and Loffredo}(2005)}]{Ga05}
\bibinfo{author}{\bibfnamefont{D.}~\bibnamefont{Garlaschelli}}
  \bibnamefont{and} \bibinfo{author}{\bibfnamefont{M.}~\bibnamefont{Loffredo}},
  \bibinfo{journal}{Physica A} \textbf{\bibinfo{volume}{355}},
  \bibinfo{pages}{138} (\bibinfo{year}{2005}).

\bibitem[{\citenamefont{Barigozzi et~al.}(2011)\citenamefont{Barigozzi,
  Fagiolo, and Mangioni}}]{Barigozzi_etal_2010physa}
\bibinfo{author}{\bibfnamefont{M.}~\bibnamefont{Barigozzi}},
  \bibinfo{author}{\bibfnamefont{G.}~\bibnamefont{Fagiolo}}, \bibnamefont{and}
  \bibinfo{author}{\bibfnamefont{G.}~\bibnamefont{Mangioni}},
  \bibinfo{journal}{Physica A: Statistical Mechanics and its Applications}
  \textbf{\bibinfo{volume}{390}}, \bibinfo{pages}{2051 }
  (\bibinfo{year}{2011}).

\bibitem[{\citenamefont{Piccardi and Tajoli}(2012)}]{Piccardi_Tajoli_2012}
\bibinfo{author}{\bibfnamefont{C.}~\bibnamefont{Piccardi}} \bibnamefont{and}
  \bibinfo{author}{\bibfnamefont{L.}~\bibnamefont{Tajoli}},
  \bibinfo{journal}{Physical Review E} \textbf{\bibinfo{volume}{85}},
  \bibinfo{pages}{066119} (\bibinfo{year}{2012}).

\bibitem[{\citenamefont{Squartini
  et~al.}(2011{\natexlab{a}})\citenamefont{Squartini, Fagiolo, and
  Garlaschelli}}]{Squartini_etal_2011a_pre}
\bibinfo{author}{\bibfnamefont{T.}~\bibnamefont{Squartini}},
  \bibinfo{author}{\bibfnamefont{G.}~\bibnamefont{Fagiolo}}, \bibnamefont{and}
  \bibinfo{author}{\bibfnamefont{D.}~\bibnamefont{Garlaschelli}},
  \bibinfo{journal}{Physical Review E} \textbf{\bibinfo{volume}{84}},
  \bibinfo{pages}{046117} (\bibinfo{year}{2011}{\natexlab{a}}).

\bibitem[{\citenamefont{Squartini
  et~al.}(2011{\natexlab{b}})\citenamefont{Squartini, Fagiolo, and
  Garlaschelli}}]{Squartini_etal_2011b_pre}
\bibinfo{author}{\bibfnamefont{T.}~\bibnamefont{Squartini}},
  \bibinfo{author}{\bibfnamefont{G.}~\bibnamefont{Fagiolo}}, \bibnamefont{and}
  \bibinfo{author}{\bibfnamefont{D.}~\bibnamefont{Garlaschelli}},
  \bibinfo{journal}{Physical Review E} \textbf{\bibinfo{volume}{84}},
  \bibinfo{pages}{046118} (\bibinfo{year}{2011}{\natexlab{b}}).

\bibitem[{\citenamefont{Fagiolo}(2010)}]{Fagiolo2010jeic}
\bibinfo{author}{\bibfnamefont{G.}~\bibnamefont{Fagiolo}},
  \bibinfo{journal}{Journal of Economic Interaction and Coordination}
  \textbf{\bibinfo{volume}{5}}, \bibinfo{pages}{1} (\bibinfo{year}{2010}).

\bibitem[{\citenamefont{Duenas and Fagiolo}(2013)}]{Duenas_Fagiolo_2013}
\bibinfo{author}{\bibfnamefont{M.}~\bibnamefont{Duenas}} \bibnamefont{and}
  \bibinfo{author}{\bibfnamefont{G.}~\bibnamefont{Fagiolo}},
  \bibinfo{journal}{Journal of Economic Interaction and Coordination}
  \textbf{\bibinfo{volume}{8}}, \bibinfo{pages}{155} (\bibinfo{year}{2013}).

\bibitem[{\citenamefont{Ward et~al.}(2013)\citenamefont{Ward, Ahlquist, and
  Rozenas}}]{Ward_etal_Gravity_Rainbow_2013}
\bibinfo{author}{\bibfnamefont{M.~D.} \bibnamefont{Ward}},
  \bibinfo{author}{\bibfnamefont{J.~S.} \bibnamefont{Ahlquist}},
  \bibnamefont{and} \bibinfo{author}{\bibfnamefont{A.}~\bibnamefont{Rozenas}},
  \bibinfo{journal}{Network Science} \textbf{\bibinfo{volume}{1}},
  \bibinfo{pages}{95} (\bibinfo{year}{2013}).

\bibitem[{\citenamefont{Rauch and Trindade}(1999)}]{Rau99}
\bibinfo{author}{\bibfnamefont{J.~E.} \bibnamefont{Rauch}} \bibnamefont{and}
  \bibinfo{author}{\bibfnamefont{V.}~\bibnamefont{Trindade}},
  \bibinfo{type}{NBER Working Papers} \bibinfo{number}{7189},
  \bibinfo{institution}{National Bureau of Economic Research, Inc}
  (\bibinfo{year}{1999}),
  \urlprefix\url{http://ideas.repec.org/p/nbr/nberwo/7189.html}.

\bibitem[{\citenamefont{Felbermayr et~al.}(2010)\citenamefont{Felbermayr, Jung,
  and Toubal}}]{Fel10}
\bibinfo{author}{\bibfnamefont{G.~J.} \bibnamefont{Felbermayr}},
  \bibinfo{author}{\bibfnamefont{B.}~\bibnamefont{Jung}}, \bibnamefont{and}
  \bibinfo{author}{\bibfnamefont{F.}~\bibnamefont{Toubal}},
  \bibinfo{journal}{Annals of Economics and Statistics}
  \textbf{\bibinfo{volume}{10}}, \bibinfo{pages}{41} (\bibinfo{year}{2010}).

\bibitem[{\citenamefont{Felbermayr and Toubal}(2012)}]{Fel12}
\bibinfo{author}{\bibfnamefont{G.~J.} \bibnamefont{Felbermayr}}
  \bibnamefont{and} \bibinfo{author}{\bibfnamefont{F.}~\bibnamefont{Toubal}},
  \bibinfo{journal}{World Development} \textbf{\bibinfo{volume}{40}},
  \bibinfo{pages}{928 } (\bibinfo{year}{2012}),
  \urlprefix\url{http://www.sciencedirect.com/science/article/pii/S0305750X11003007}.

\bibitem[{\citenamefont{Ozden et~al.}(2011)\citenamefont{Ozden, Parsons,
  Schiff, and Walmsley}}]{Ozden_etal_data_2011}
\bibinfo{author}{\bibfnamefont{C.}~\bibnamefont{Ozden}},
  \bibinfo{author}{\bibfnamefont{C.~R.} \bibnamefont{Parsons}},
  \bibinfo{author}{\bibfnamefont{M.}~\bibnamefont{Schiff}}, \bibnamefont{and}
  \bibinfo{author}{\bibfnamefont{T.~L.} \bibnamefont{Walmsley}},
  \bibinfo{journal}{World Bank Economic Review} \textbf{\bibinfo{volume}{25}},
  \bibinfo{pages}{12} (\bibinfo{year}{2011}).

\bibitem[{\citenamefont{Gleditsch}(2002)}]{GledData2002}
\bibinfo{author}{\bibfnamefont{K.}~\bibnamefont{Gleditsch}},
  \bibinfo{journal}{Journal of Conflict Resolution}
  \textbf{\bibinfo{volume}{46}}, \bibinfo{pages}{712} (\bibinfo{year}{2002}).

\bibitem[{\citenamefont{Fagiolo}(2006)}]{Fagiolo2006EcoBull}
\bibinfo{author}{\bibfnamefont{G.}~\bibnamefont{Fagiolo}},
  \bibinfo{journal}{Economics Bulletin} \textbf{\bibinfo{volume}{3}},
  \bibinfo{pages}{1} (\bibinfo{year}{2006}).

\bibitem[{\citenamefont{van Bergeijk and Brakman}(2010)}]{GravityBook}
\bibinfo{editor}{\bibfnamefont{P.}~\bibnamefont{van Bergeijk}}
  \bibnamefont{and} \bibinfo{editor}{\bibfnamefont{S.}~\bibnamefont{Brakman}},
  eds., \emph{\bibinfo{title}{The Gravity Model in International Trade}}
  (\bibinfo{publisher}{Cambridge University Press, Cambridge},
  \bibinfo{year}{2010}).

\bibitem[{\citenamefont{Lewer and Van~den Berg}(2008)}]{Lewer_2008}
\bibinfo{author}{\bibfnamefont{J.~J.} \bibnamefont{Lewer}} \bibnamefont{and}
  \bibinfo{author}{\bibfnamefont{H.}~\bibnamefont{Van~den Berg}},
  \bibinfo{journal}{Economics Letters} \textbf{\bibinfo{volume}{99}},
  \bibinfo{pages}{164} (\bibinfo{year}{2008}).

\bibitem[{\citenamefont{Saramaki et~al.}(2007)\citenamefont{Saramaki,
  Kivel\"{a}, Onnela, Kaski, and Kert\'{e}sz}}]{Saramaki2006}
\bibinfo{author}{\bibfnamefont{J.}~\bibnamefont{Saramaki}},
  \bibinfo{author}{\bibfnamefont{M.}~\bibnamefont{Kivel\"{a}}},
  \bibinfo{author}{\bibfnamefont{J.}~\bibnamefont{Onnela}},
  \bibinfo{author}{\bibfnamefont{K.}~\bibnamefont{Kaski}}, \bibnamefont{and}
  \bibinfo{author}{\bibfnamefont{J.}~\bibnamefont{Kert\'{e}sz}},
  \bibinfo{journal}{Physical Review E} \textbf{\bibinfo{volume}{75}},
  \bibinfo{pages}{027105} (\bibinfo{year}{2007}).

\bibitem[{\citenamefont{Fagiolo}(2007)}]{Fa07}
\bibinfo{author}{\bibfnamefont{G.}~\bibnamefont{Fagiolo}},
  \bibinfo{journal}{Physical Review E} \textbf{\bibinfo{volume}{76}},
  \bibinfo{pages}{026107} (\bibinfo{year}{2007}).

\bibitem[{\citenamefont{Kleinberg}(1999)}]{Kleinberg_1999}
\bibinfo{author}{\bibfnamefont{J.}~\bibnamefont{Kleinberg}},
  \bibinfo{journal}{Journal of the ACM} \textbf{\bibinfo{volume}{465}},
  \bibinfo{pages}{604} (\bibinfo{year}{1999}).

\end{thebibliography}
\end{document}